\documentclass[sigconf]{acmart}

\def\BibTeX{{\rm B\kern-.05em{\sc i\kern-.025em b}\kern-.08emT\kern-.1667em\lower.7ex\hbox{E}\kern-.125emX}}
    
\usepackage{acronym}
\newacro{sme}[SME]{Small and Medium-sized Enterprise}
\newacro{it}[IT]{Information Technology}
\newacro{ot}[OT]{Operation Technology}
\newacro{cots}[COTS]{Commercial Off-The-Shelf}
\newacro{iot}[IoT]{Internet of Things}
\newacro{iiot}[IIoT]{Industrial Internet of Things}
\newacro{plc}[PLC]{Programmable Logic Controller}
\newacro{cps}[CPS]{Cyber-Physical System}
\newacro{cpps}[CPPS]{Cyber-Physical Production System}
\newacro{ids}[IDS]{Intrusion Detection System}
\newacro{svm}[\textit{SVM}]{\textit{Support Vector Machine}}
\newacro{wsn}[WSN]{Wireless Sensor Network}
\newacro{darpa}[DARPA]{Defense Advanced Research Projects Agency}
\newacro{kdd}[KDD]{Knowledge Discovery in Databases}
\newacro{scada}[SCADA]{Supervisory Control And Data Acquisition}
\newacro{dpi}[DPI]{Deep Packet Inspection}
\newacro{rnn}[RNN]{Recurrent Neural Network}
\newacro{s317}[\textit{S317}]{\textit{SUTD Security Showdown (S3) 2017}}
\newacro{dmz}[DMZ]{De-Militarized Zone}
\newacro{svm}[\textit{SVM}]{\textit{Support Vector Machine}}
\newacro{ocsvm}[\textit{OCSVM}]{\textit{One-Class Support Vector Machine}}
\newacro{lstm}[\textit{LSTM}]{\textit{Long Short-Term Memory}}
\newacro{iiot}[IIoT]{Industrial Internet of Things}
\newacro{opcua}[\textit{OPC UA}]{\textit{Object Linking and Embedding for Process Control Unified Architecture}}
\newacro{swat}[\textit{SWaT}]{\textit{Secure Water Treatment}}
\newacro{pca}[\textit{PCA}]{\textit{Principal Component Analysis}}
\newacro{sssp}[SSSP]{Single Stage Single Point}
\newacro{ssmp}[SSMP]{Single Stage Multi Point}
\newacro{mssp}[MSSP]{Multi Stage Single Point}
\newacro{msmp}[MSMP]{Multi Stage Multi Point}
\newacro{mtu}[MTU]{Master Terminal Unit}
\newacro{uf}[UF]{Ultra Filtration}
\newacro{uv}[UV]{Ultraviolet}    
\newacro{ro}[RO]{Reverse Osmosis} 		
\newacro{rbf}[RBF]{Radial Basis Function} 

\usepackage{amsmath}	
\usepackage{mathtools}

\usepackage[english]{babel}

\usepackage{multirow}

\usepackage{numprint}
\npdecimalsign{.}

\DeclarePairedDelimiter\norm{\lVert}{\rVert}%
    
\setcopyright{acmlicensed}
\copyrightyear{2019}
\acmYear{2019}
\acmConference[CECC 2019]{Central European Cybersecurity Conference}{November 14--15, 2019}{Munich, Germany}
\acmBooktitle{Central European Cybersecurity Conference (CECC 2019), November 14--15, 2019, Munich, Germany}
\acmPrice{15.00}
\acmDOI{10.1145/3360664.3360669}
\acmISBN{978-1-4503-7296-1/19/11}

\begin{document}

%
\title[Security in Process]{Security in Process:\\Detecting Attacks in Industrial Process Data}

%
\author{Simon D. Duque Anton}
\email{simon.duque_anton@dfki.de}
\orcid{0000-0003-4005-9165}
\affiliation{%
  \institution{German Research Center for AI\\Intelligent Networks Research Group}
  \streetaddress{Trippstadter Str. 122}
  \city{Kaiserslautern}
  \country{Germany}
  \postcode{67633}
}

\author{Anna Pia Lohfink}
\email{lohfink@cs.uni-kl.de}
\affiliation{
  \institution{University of Kaiserslautern\\Department of Computer Science}
  \streetaddress{Trippstadter Str. 122}
  \city{Kaiserslautern}
  \country{Germany}
  \postcode{67633}
}

\author{Christoph Garth}
\email{garth@cs.uni-kl.de}
\affiliation{
  \institution{University of Kaiserslautern\\Department of Computer Science}
  \streetaddress{Trippstadter Str. 122}
  \city{Kaiserslautern}
  \country{Germany}
  \postcode{67633}
}

\author{Hans Dieter Schotten}
\email{hans_dieter.schotten@dfki.de}
\affiliation{
  \institution{German Research Center for AI\\Intelligent Networks Research Group}
  \streetaddress{Trippstadter Str. 122}
  \city{Kaiserslautern}
  \country{Germany}
  \postcode{67633}
}

%
\renewcommand{\shortauthors}{S. D. Duque Anton et al.}

%
\begin{abstract}
Due to the fourth industrial revolution,
industrial applications make use of the progress in communication and embedded devices.
This allows industrial users to increase efficiency and manageability while reducing cost and effort.
Furthermore,
the fourth industrial revolution,
creating the so-called \textit{Industry 4.0},
opens a variety of novel use and business cases in the industrial environment.
However,
this progress comes at the cost of an enlarged attack surface of industrial companies.
Operational networks that have previously been phyiscally separated from public networks are now connected in order to make use of new communication capabilites.
This motivates the need for industrial intrusion detection solutions that are compatible to the long-term operation machines in industry as well as the heterogeneous and fast-changing networks.
In this work,
process data is analysed.
The data is created and monitored on real-world hardware.
After a set up phase,
attacks are introduced into the systems that influence the process behaviour.
A time series-based anomaly detection approach,
the \textit{Matrix Profiles},
are adapted to the specific needs and applied to the intrusion detection.
The results indicate an applicability of these methods to detect attacks in the process behaviour.
Furthermore,
they are easily integrated into existing process environments.
Additionally,
one-class classifiers \textit{One-Class Support Vector Machines} and \textit{Isolation Forest} are applied to the data without a notion of timing.
While \textit{Matrix Profiles} perform well in terms of creating and visualising results,
the one-class classifiers perform poorly.
\end{abstract}

%
%


%

%
\maketitle
\selectlanguage{english}
\section{Introduction}
\label{sec:intro}
Attacks on industrial enterprises have increased over the last two decades,
in frequency as well as in impact~\cite{Duque_Anton.2017a}.
A number of differences between home and office \ac{it} networks makes classic \ac{it} security solutions only partly applicable to the industrial environment.
Heterogeneous networks,
consisting of physically distributed devices with operation times of several decades make updates and fixes difficult.
Communication protocols such as \textit{Modbus/TCP}~\cite{Modbus.2012, ModbusIDA.2006} are employed that do not contain means of authentication or encryption.
That means any attacker having obtained access to the communication network is capable of reading and injecting messages.
For two historic reasons,
industrial,
\ac{scada} or \ac{ot} networks are not secured in a fashion deemed appropriate for home and office \ac{it}~\cite{Igure.2006}.
First,
\ac{ot} networks are supposed to be physically separated from \ac{it} networks.
Second,
\ac{ot} networks and their connected devices are considered to be highly application specific,
thus making propagation and exploitation by an attacker difficult.
Both reasons do not hold true anymore.
\ac{cots} products in the industrial area such as \acp{plc} make set up,
maintenance and operation of industrial applications much easier due to common interfaces and programming libraries.
Those,
however,
make it easier for an attacker to prepare for exploitation as well.
Furthermore,
the fourth industrial revolution introduces an abundance of novel use and business cases~\cite{3gpp2017}.
Most of them rely on the communication and computation capabilities of \ac{iot} and \ac{iiot} devices.
This breaks the physical separation of networks,
creating access routes to \ac{ot} networks.
Even if no such access is possible,
attackers have successfully managed to move laterally to the \ac{ot} networks in the past after breaking the \ac{it} network perimeter~\cite{Duque_Anton.2019c}.
As a result,
efficient intrusion detection in industrial networks is crucial for the secure and sound operation of industrial applications.
However, 
several attacks are capable of masking sensor outputs once the attacker gains access to the device under attack.
Thus, 
correlation of several sensor values,
preferably over separate channels,
can be employed to detect attacks.\\
In this work,
machine learning- as well as time series-based methods for anomaly detection are applied to one set of process data of a real-world industrial process,
containing several sensors and actuators,
controlled by six \acp{plc}.
As the data was created and captured in a experimental environment,
ground truth about the attacks is available,
i.e. there are labels and the assurance that the data is labelled correctly.
The contribution of this paper is twofold:
\begin{itemize}
\item The feasibility of multi-sensor anomaly detection based on machine learning is evaluated
\item a time series motif discovery algorithm is extended for anomaly detection
\end{itemize}
The remainder of this work is structured as follows.
In Section~\ref{sec:sota},
an overview of the state of the art is provided.
The data set analysed in this work are presented in Section~\ref{sec:data},
the algorithms used to evaluate the data are introduced in Section~\ref{sec:algos}.
In Section~\ref{sec:eval},
they are evaluated.
This work is concluded in Section~\ref{sec:concl}.

\section{Related Work}
\label{sec:sota}
Due to the increasing relevance of industrial intrusion detection,
it is a widely regarded topic in the research community.
\textit{Schneider and B\"{o}ttinger} use \textit{autoencoders} in order to detect real attacks in an industrial data set in an unsupervised fashion~\cite{Schneider.2018}.
A framework for assessing the impact of cyber attacks in production environments is presented by \textit{Giehl et al.}~\cite{Giehl.2019}.
They evaluate their approach on the \ac{s317} data set provided by the \textit{iTrust,
Centre for Research in Cyber Security,
Singapore University of Technology and Design}~\cite{iTrust.2017}.
The attack data has been generated by a contest to affect the process environment with real attacks.
\textit{Goh et al.} use \acp{rnn} to detect attacks on a \acp{cps} in a data set by the same institution~\cite{Goh.2017}.
They evaluate the \ac{swat} data set~\cite{iTrust.2018},
the same data set that has been evaluated in the course of this work.
Creating realistic data sets for industrial intrusion detection is crucial for the development and testing of intrusion detection systems.
A special kind of neural networks,
\ac{lstm},
is used by \textit{Feng et al.}~\cite{Feng.2017}.
They employ a multi-level approach in order to detect attacks in a gas pipeline data set.
\textit{Knapp and Langill} present approaches to secure industrial networks~\cite{Knapp.2014}.
The detection of intrusions in power system networks is discussed by \textit{Yang et al.}~\cite{Yang.2014}.
\textit{Larkin et al.} present a summary about the evolution of \ac{scada} security systems~\cite{Larkin.2014}.
\ac{ocsvm} as a machine learning algorithm to detect novel and unknown attacks is presented by \textit{Maglaras and Jiang}~\cite{Maglaras.2014}.
Approaches to detect attacks in \textit{Modbus} data with the help of signatures is presented by \textit{Gao and Morris}~\cite{Gao.2014}.
The security of future industrial applications with the integration of the \ac{iiot} is addressed by \textit{Plaga et al.}~\cite{Plaga.2019, Plaga.2018}.

\section{Data Set}
\label{sec:data}
\begin{figure}
  \includegraphics[width=\linewidth]{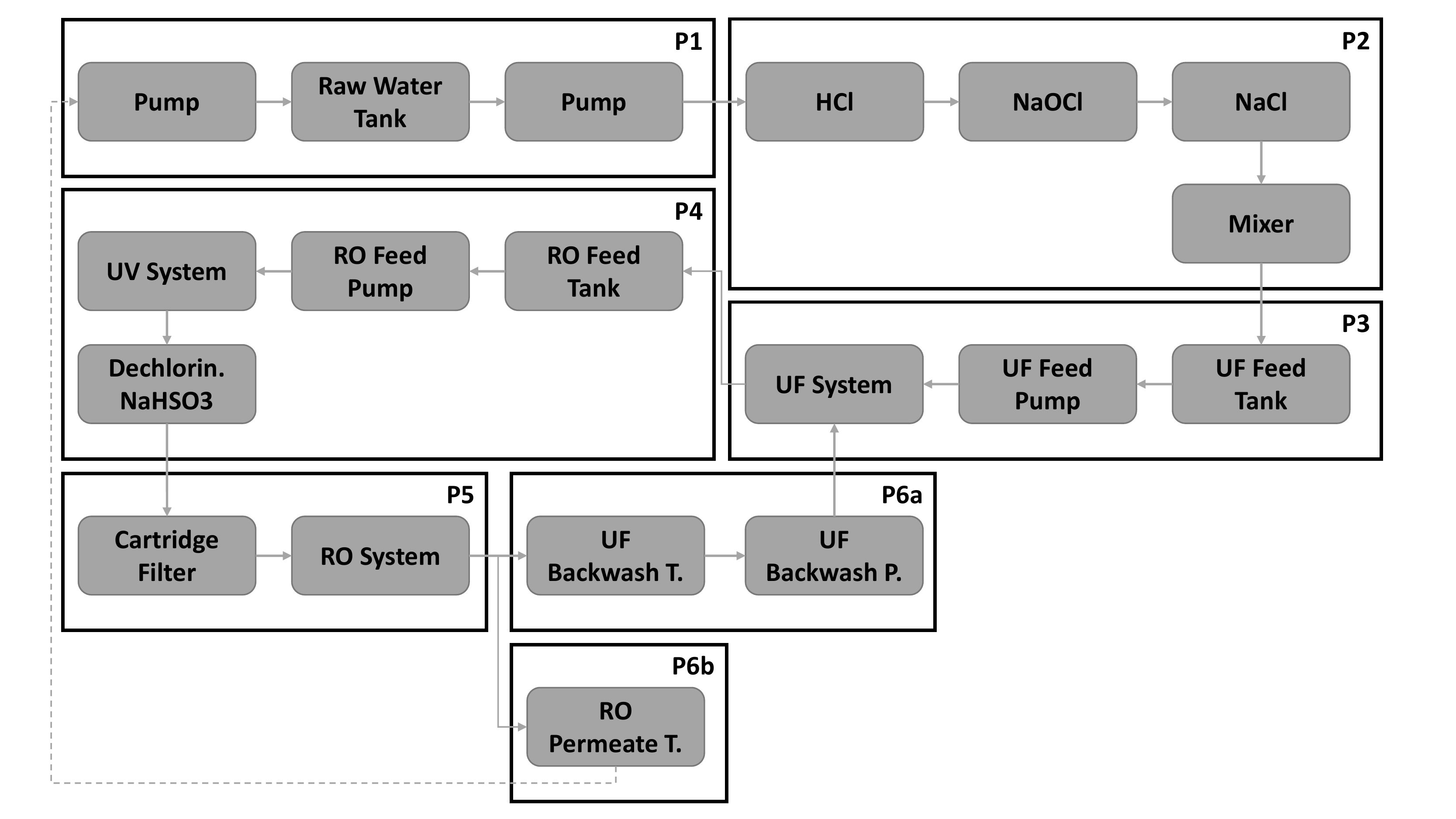}
  \caption{Relation of Sub-Processes}
  \label{fig:process_order}
\end{figure}

\begin{figure}
  \includegraphics[width=\linewidth]{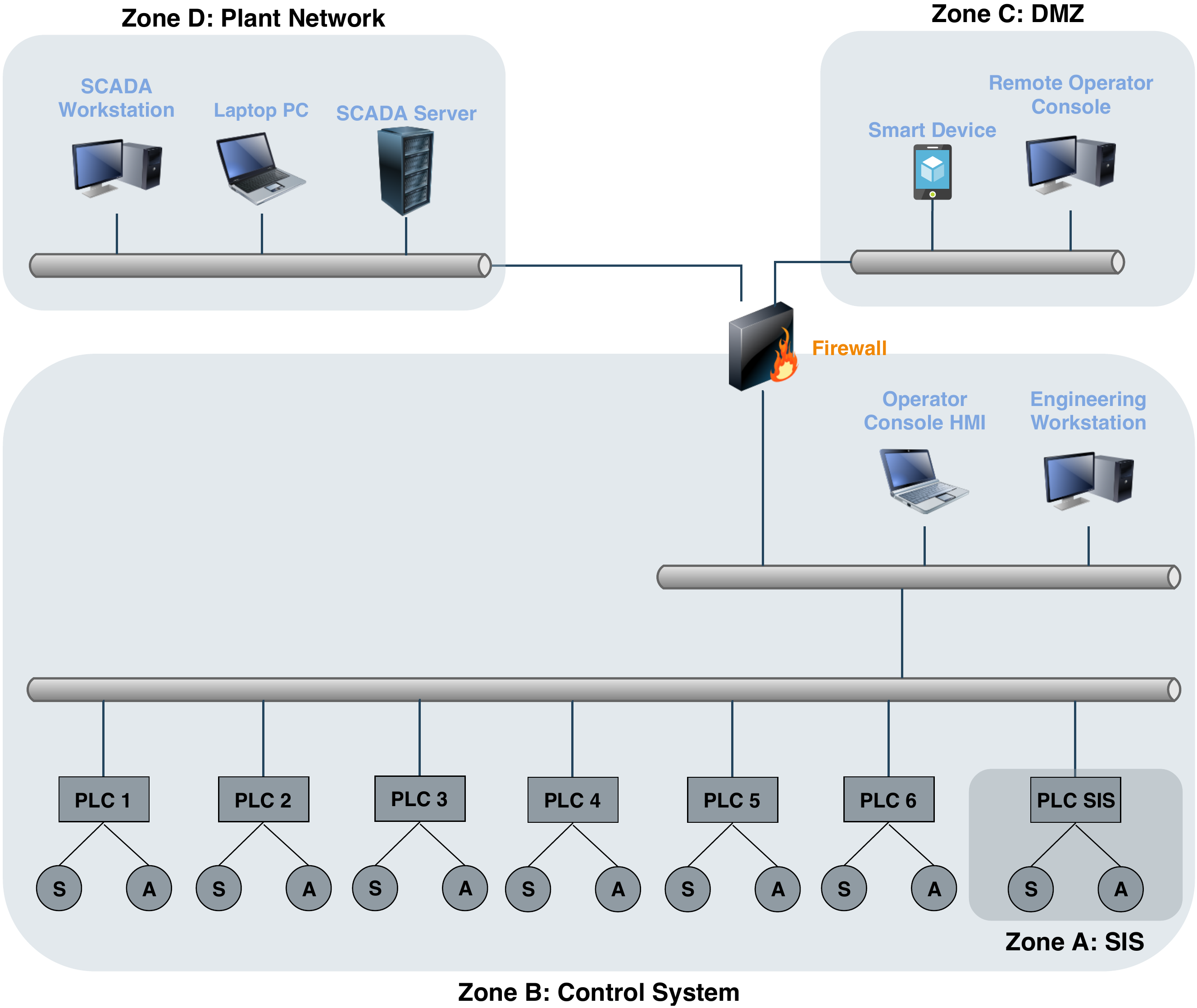}
  \caption{Schematic Overview of the Process Environment}
  \label{fig:nw_structure}
\end{figure}

An industrial data set is evaluated in this work.
It is presented by the \textit{iTrust,
Centre for Research in Cyber Security,
Singapore University of Technology and Design} and is called \ac{swat}~\cite{Goh.2016, iTrust.2018}.
It is taken in an industrial environment with a real-world underlying process into which attacks have been introduced.
The data set is gathered from a water processing facility that was set up in a laboratory context.
It consists of the following six sub-processes:
\begin{itemize}
\item \textit{P1}: Raw water storage
\item \textit{P2}: Pre-treatment
\item \textit{P3}: Membrane \ac{uf}
\item \textit{P4}: Dechlorination by \ac{uv} lamps
\item \textit{P5}: \ac{ro}
\item \textit{P6}: Disposal
\end{itemize}
The relations of and transitions between the sub-processes are depicted in Figure~\ref{fig:process_order}.
First,
the water to be processed is stored,
followed by a pre-treatment with different chemicals.
After that, 
\ac{uf} is applied,
followed by an \ac{uv} process.
It is then pumped to an \ac{ro} process.
Depending on the level of cleanliness,
it is either stored in a clean water reservoir or fed back to the \ac{uf} process.
The environment in which the \acp{plc} controlling the sub-processes are set up is depicted in Figure~\ref{fig:nw_structure}.
The \acp{plc} in this figure correspond to the sub-processes with the respective number.
In total,
they control 51 sensors and actuators.
An exhaustive list including description can be found in the work of \textit{Goh et al.}~\cite{Goh.2016}.
They distinguish between four types of attacks,
describing the type and listing the number of occurrences in the data:
\begin{itemize}
\item \textit{\ac{sssp}}: Single stage attack on one point in the process, 26 instances in the data set
\item \textit{\ac{ssmp}}: Single stage attack on multiple points in the process, 4  instances in the data set
\item \textit{\ac{mssp}}: Multi stage attack on one point in the process, 2 instances in the data set
\item \textit{\ac{msmp}}: Multi stage attack on multiple points in the process, 4 instances in the data set
\end{itemize}

\section{Algorithms Used}
\label{sec:algos}
In this work,  
three different algorithms have been used to determine attacks in the data set:
\acp{ocsvm},
\textit{Isolation Forests} and \textit{Matrix Profiles}.
\acp{ocsvm} and \textit{Isolation Forests} have been used to analyse the data sets,
as described in Section~\ref{sec:data},
on packet basis while \textit{Matrix Profiles} have been used to perform a time series analysis of the data sets.

\subsection{\textit{One-Class Support Vector Machines}}
\acp{ocsvm} are a one-class classifier that is trained with one class in order to determine whether elements of the test case do or do not belong to that class.
It is well suited for applications where one class is predominant,
such as anomaly detection.
\acp{svm} are a large-margin classifier.
They were introduced by \textit{Boser et al.} in 1992~\cite{Boser.1992}.
Elements in an $n$-dimensional data space are separated by an $n+1$-dimensional hyperplane.
The elements are described as tuples as shown in (\ref{eq:svm_data})~\cite{Cortes.1995}.
\begin{equation}
\label{eq:svm_data}
\begin{split}
(x_{i}, y_{i}), i = 1, ..., m, y \in \{-1, 1\}
\end{split}
\end{equation}
$x$ is a vector of dimension $n$,
$y$ is an attribution indicating affiliation to one  of two classes.
In classic \ac{svm},
both classes are present in the training data,
leading to a hyperplane separating both classes so that the distance of every element to the hyperplane is maximal.
In one-class classification according to \textit{Sch\"{o}lkopf et al.}~\cite{Schoelkopf.2000},
the distances of data points from the hyperplane are maximised.
A quadratic minimisation function (\ref{eq:min_svm}) minimises the area surrounded by the hyperplane according to $\nu$.
\begin{equation}
\label{eq:min_svm}
\begin{split}
\min_{\omega, \xi_{i}, \rho} \frac{1}{2} \norm{\omega}^{2} + \frac{1}{\nu n} \sum_{i=1}^{n} \xi_{i} - \rho \\
\text{with}\\
(\omega \times \phi(x_{i})) \geq \rho - \xi_{i} \forall i \in \{1, ..., n\} \\ 
\xi_{i} \geq 0 \forall i \in \{1, ..., n\}
\end{split}
\end{equation}
$\nu \in (0, 1)$ characterises the fraction of outliers that are tolerated as well as the training size.
$\omega$ and $\rho$ are used to describe the hyperplane.
After training,
the decision function (\ref{eq:dec_func}) decides whether or not an element is part of the trained class or not.
\begin{equation}
\label{eq:dec_func}
\begin{split}
f(x) = \text{sgn}((\omega \times \phi(x_{i})-\rho)
\end{split}
\end{equation}

\subsection{\textit{Isolation Forest}}
Similar to \ac{ocsvm},
\textit{Isolation Forest} is a one-class classifier~\cite{Liu.2008}.
The goal is to isolate anomalous data points in a data set.
This is done by training an ensemble of decision trees on a data set and then considering the path length until convergence as a metric for isolation.
Densely populated areas converge quickly,
while isolated areas take longer to converge.
A data sample is described by (\ref{eq:data_if}).
\begin{equation}\label{eq:data_if}
\begin{split}
X = \{x_1, ... , x_n\}
\end{split}
\end{equation}
A tree is created by selecting random attributes $q$ as well as a split value $p$ until either:
\begin{itemize}
\item the height limit of the tree is reached,
\item $|X| = 1$ or
\item all data in the tree is of the same value.
\end{itemize}
The average path length is calculated and derived from this,
the path length of each sample is calculated and an anomaly score can be derived. 
In sorting the path lengths in ascending order,
anomalies are found at the top of the list.

\subsection{\textit{Matrix Profiles}}
\textit{Matrix Profiles} were developed in 2016 by \textit{Yeh et al.}~\cite{Yeh.2016a} as an algorithm for motif discovery.
A time series data set was split into sequences of length $m$.
The distance of each sequence starting at a point in the data set from each other sequence is then calculated in a sliding window fashion,
e.g. with the z-normalised distance (\ref{eq:z_norm_dist}).
\begin{equation}\label{eq:z_norm_dist}
\begin{split}
d(x,y) = \sqrt{\sum_{i=1}^{m}{(\hat{x}_{i} - \hat{y}_{i})}^2} \\
\hat{x}_{i} = \frac{x_{i} - \mu_{x}}{\sigma_{x}},\quad \hat{y}_{i} = \frac{y_{i} - \mu_{y}}{\sigma_{y}}
\end{split}
\end{equation}
After applying \textit{Pearson's Correlation Coefficient}~\cite{Benesty.2009} (\ref{eq:pearson})
\begin{equation}\label{eq:pearson}
\begin{split}
corr(x,y) &= \frac{E((x - \mu_x)(y-\mu_y))}{\sigma_x \sigma_y} \\
& = \frac{\sum^{m}_{i=1}x_i y_i - m \mu_x \mu_y}{m \sigma_x \sigma_y},
\end{split}
\end{equation}
where
\begin{equation}\label{eq:mu}
\begin{split}
\mu_x = \frac{\sum_{i=1}^{m} x_i}{m}, \quad \mu_y = \frac{\sum_{i=1}^{m} y_i}{m}
\end{split}
\end{equation}
and
\begin{equation}\label{eq:sigma}
\begin{split}
\sigma_{x}^{2} = \frac{\sum_{i=1}^{m} x_{i}^{2}}{m} - \mu_{x}^{2}, \quad \sigma_{y}^{2} = \frac{\sum_{i=1}^{m} y_{i}^{2}}{m} - \mu_{y}^{2}.
\end{split}
\end{equation}
The Euclidean distance relates as in (\ref{eq:relation})~\cite{Mueen.2010},
\begin{equation}\label{eq:relation}
\begin{split}
d(x,y) = \sqrt{2m(1-\text{corr}(x,y))}
\end{split}
\end{equation}
the resulting metric for distance calculation is described in (\ref{eq:working_formular_dist}).
\begin{equation}\label{eq:working_formular_dist}
\begin{split}
d(x, y) = \sqrt{2m\bigg(1-\frac{\sum_{i=1}^{m} x_{i} y_{i} - m \mu_{x} \mu_{y}}{m \sigma_{x} \sigma_{y}}\bigg)}
\end{split}
\end{equation}
$x$ and $y$  are time series,
$\mu$ is the respective mean and $\sigma$ the respective standard deviation.
The minimal distances are derived and stored in a matrix,
hence the name.
A high minimal distance indicates an outlier,
as no sequence in the time series is similar.
Correspondingly,
a low minimal distance indicates at least one similar sequence in the series.

\section{Evaluation}
\label{sec:eval}
In this section,
the application of the algorithms introduced in Section~\ref{sec:algos} on the data set as presented in Section~\ref{sec:data} is evaluated.
All algorithms employed in the course of this work only need to be trained on normal data.
This is based on the assumption that in real applications,
anomalous data is sparse.
Furthermore,
since the anomalies of interest in this work are due to attacks,
they might be unique in their characteristic and thus hard to train a priori.
Having one-class classifiers or predictors is justified by the reality of industrial environments that data from normal operation of productive systems is available in abundance while anomalous data is hardly present.

\subsection{\textit{Matrix Profiles}}
\begin{figure}[!ht]
\centering
\includegraphics[width=\linewidth]{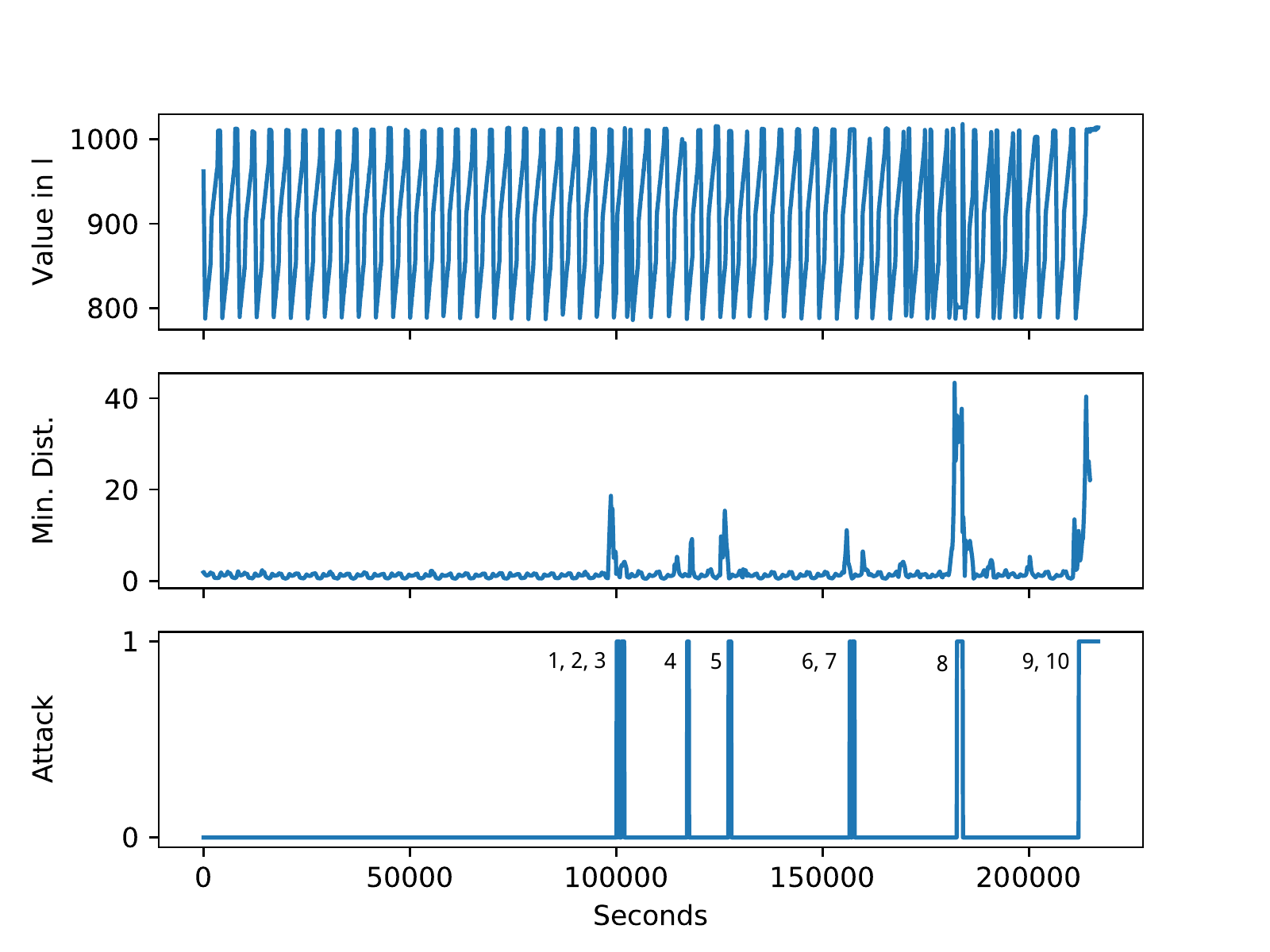}
\caption{\textit{Matrix Profile} of LIT-301}
\label{fig:LIT-301}
\end{figure}

\begin{figure}[!ht]
\centering
\includegraphics[width=\linewidth]{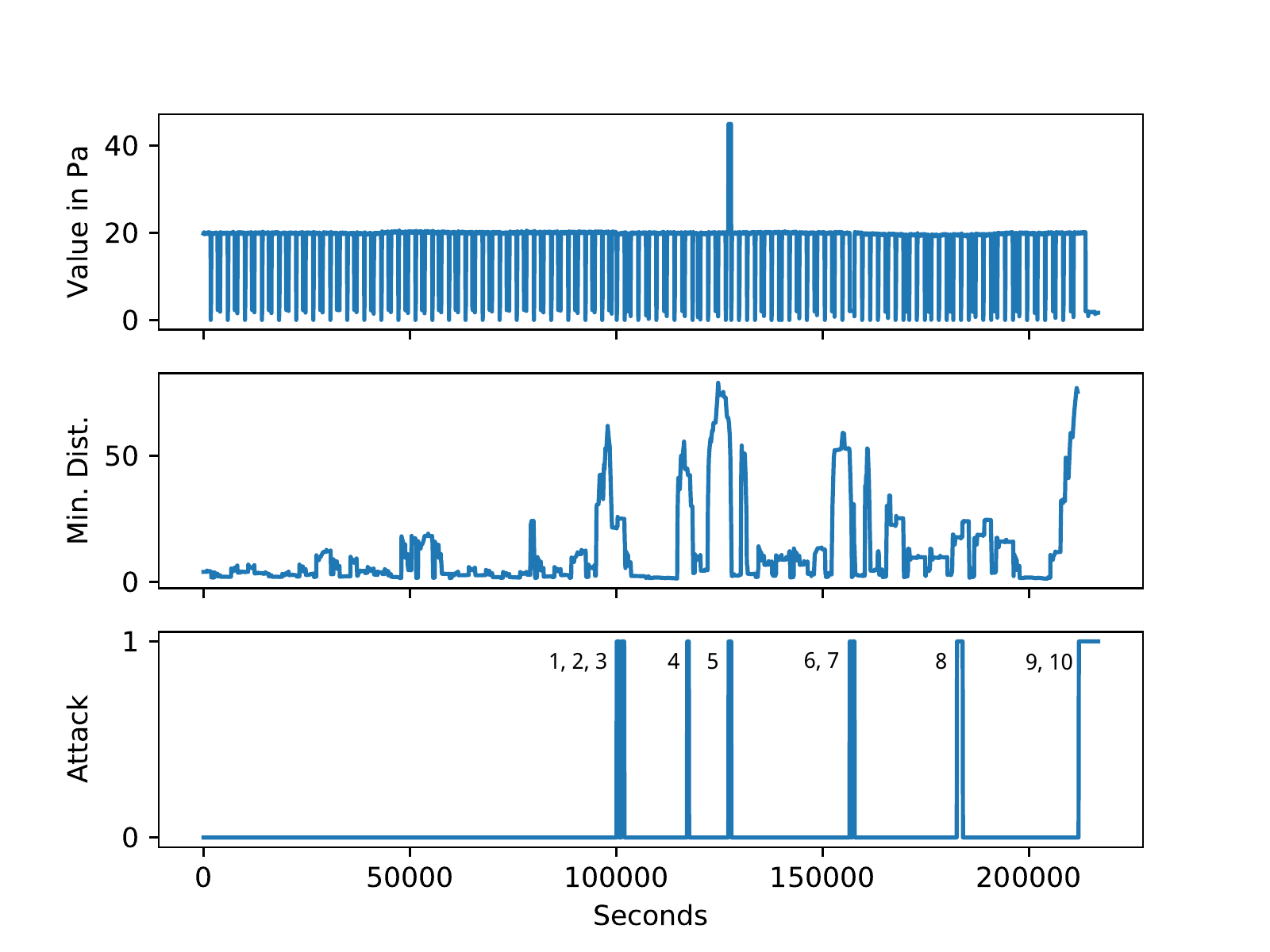}
\caption{\textit{Matrix Profile} of DPIT-301}
\label{fig:DPIT-301}
\end{figure}

\begin{figure}[!ht]
\centering
\includegraphics[width=\linewidth]{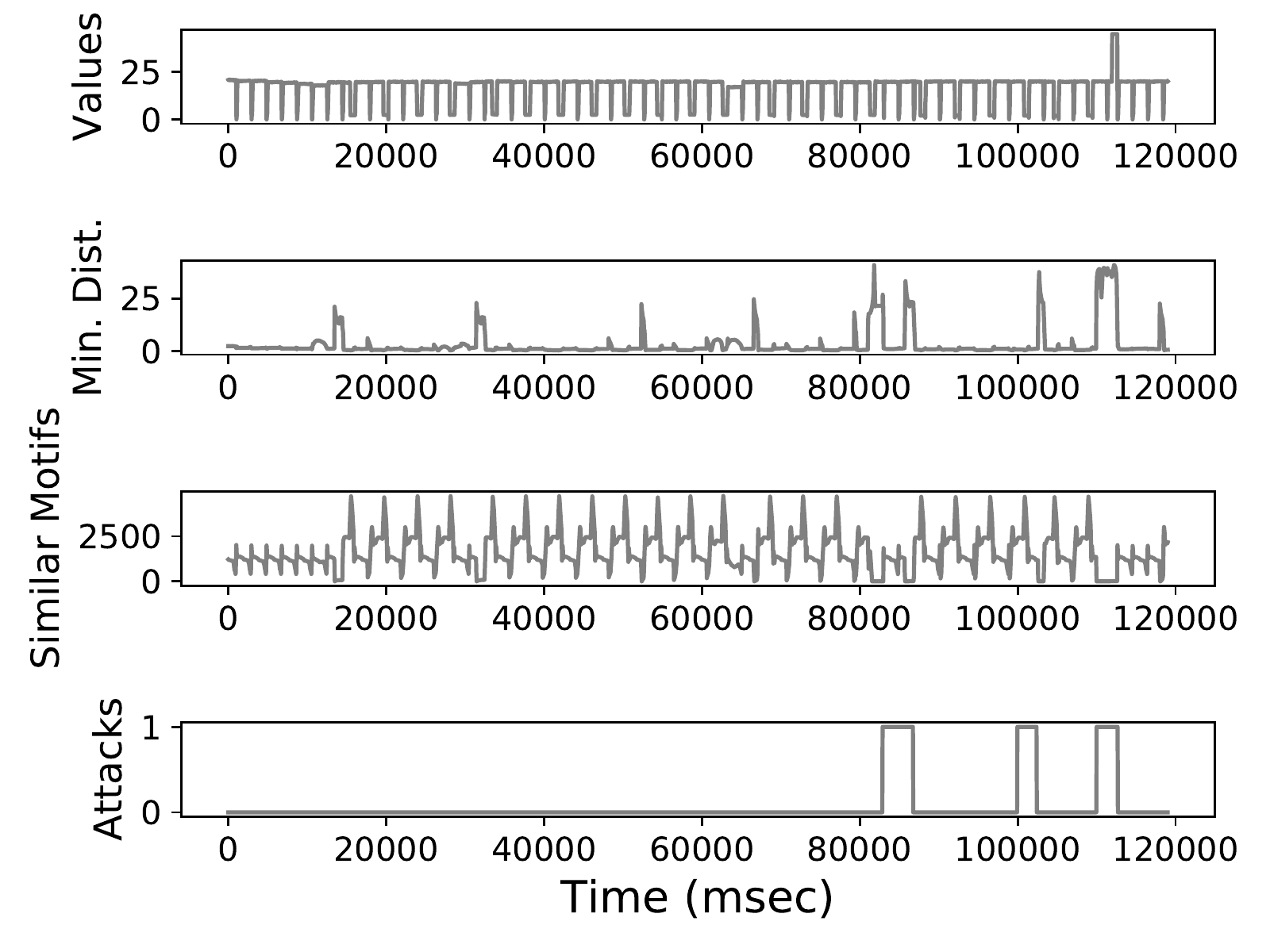}
\caption{\textit{Matrix Profile} and Similiarity Information of DPIT-301}
\label{fig:DPIT-301_new_long}
\end{figure}

\begin{figure}[!ht]
\centering
\includegraphics[width=\linewidth]{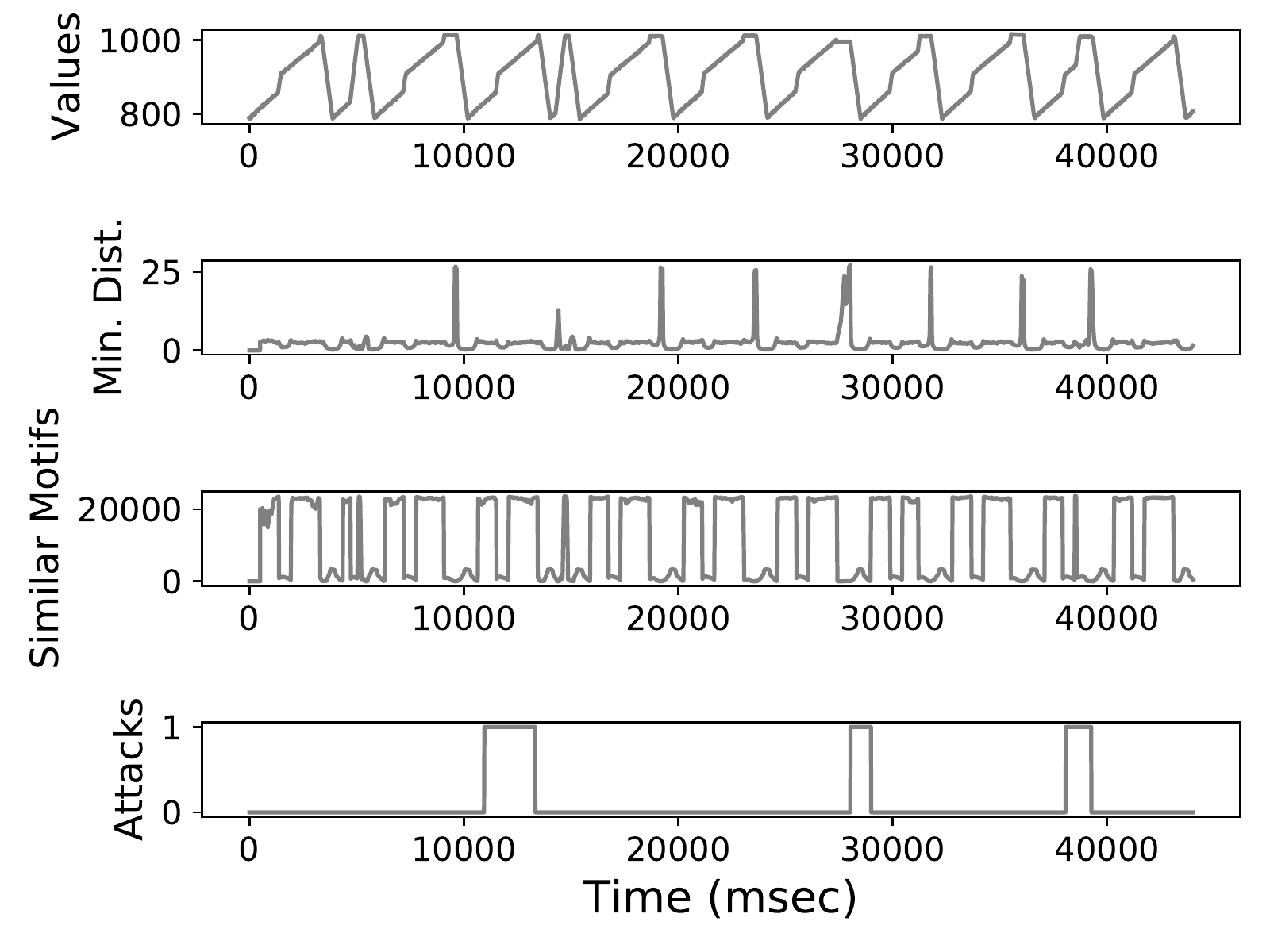}
\caption{\textit{Matrix Profile} and Similiarity Information of LIT-301 in a Smaller Interval}
\label{fig:LIT-301_new}
\end{figure}

\begin{figure}[!ht]
\centering
\includegraphics[width=\linewidth]{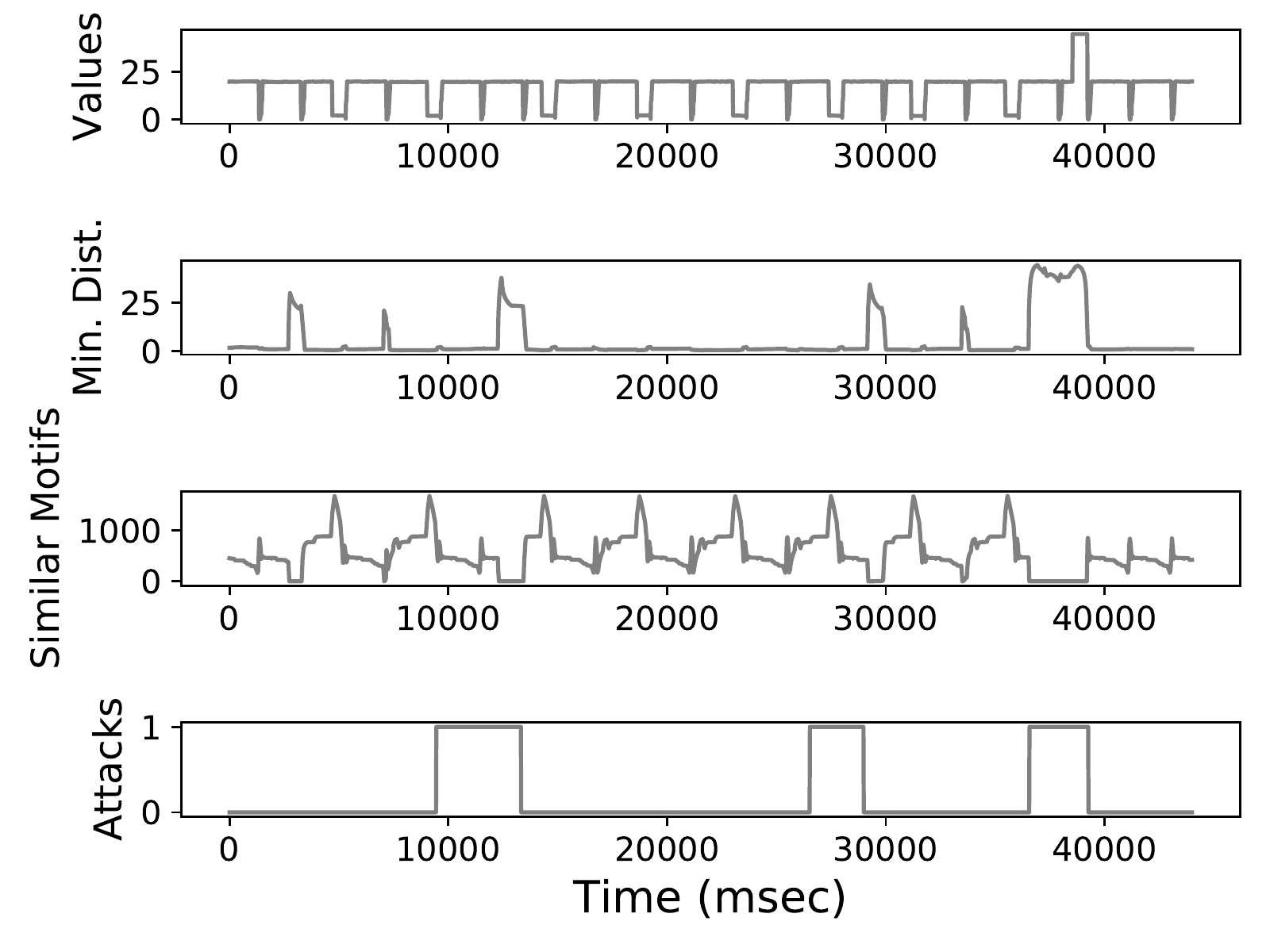}
\caption{\textit{Matrix Profile} and Similiarity Information of DPIT-301 in a Smaller Interval}
\label{fig:DPIT-301_new}
\end{figure}

In this work,
\textit{Matrix Profiles} are applied to the data set in order to determine thresholds of the minimal distance as well as create an additional metric:
the number of similar instances.
Preliminary work shows the effectiveness of \textit{Matrix Profiles} for the detection of attacks in process behaviour~\cite{Duque_Anton.2019f}.
The reference implementation of \textit{Matrix Profiles} was used and customised for the application purposes.
In Figure~\ref{fig:LIT-301},
the level of raw water tank \textit{LIT-301} is shown.
In this figure,
about \numprint{100000} time steps of normal behaviour are preceding about \numprint{150000} time steps of behaviour during which ten attacks occur.
The attacks are indicated and numbered in the bottom row by a boolean value 1.
Additionally,
the value of the sensor is shown in the first row and the minimal distance as calculated with the \textit{Matrix Profile} is shown in the second row.
It can be seen that all attacks can be detected by the increase in minimal distance as calculated with the \textit{Matrix Profile}.
The hyper-parameter $m$ was set to \numprint{500}.
Additionally,
a sensor measuring differential pressure in the backwash-process \textit{DPIT-301} was analysed in the same fashion,
shown in Figure~\ref{fig:DPIT-301}.
Its behaviour is more rugged than the behaviour of \textit{LIT-301},
the hyper-parameter $m$ is set to 2000,
the period of the sensor value.
Attacks 1, 2, and 3 are detected in this process as well,
similar to attacks 4 and 5, 6 and 7,
as well as 9 and 10.
Attack 8 is lower than noise in the minimal distance and thus not detectable.
Since \textit{Matrix Profiles} have been shown to perform well for such time series~\cite{Duque_Anton.2019e},
especially in conjunction of different aspects of the process,
they are evaluated for an extension in this work.
Since \textit{Matrix Profiles} calculate the minimal distance,
an attack that occurs twice and has the same characteristic each time is not detected as an attack anymore,
since it is already known behaviour.
This means in praxis,
an operator would have to detect every attack on the first try;
in historical analysis, 
even this would not be possible.
In order to create a more versatile approach,
the instances of a motif are counted.
An epsilon value is set to compare the current motif to all other motifs,
all motifs whose distance is smaller than the epsilon value are added to a list.
In doing so,
the number of similar motifs can be extracted.
Even if an attack occurs more than once and the masking of attack and normal behaviour results in the same pattern,
the number of similar motifs is small,
indicating a rare behaviour despite a low minimal distance.
In addition to the standard \textit{Matrix Profiles},
an extension has been implemented:
In Figure~\ref{fig:DPIT-301_new_long},
this information is added in the line \textit{Similar Values}.
This figure depicts data consisting of \numprint{85000} instances from non-malicious data as well as about \numprint{60000} instances of malicious data in which three blocks of attacks are expected.
Attacks are indicated by a one in the corresponding line.
It can be seen that the attack blocks are responsible for the highest minimal distances,
as expected and already shown in Figure~\ref{fig:DPIT-301}.
Additionally,
the similarities correspond to the \textit{Matrix Profiles}. 
As a threshold for similarities,
20 was set.
Even though there are two regions,
around \numprint{18000} and \numprint{32000} milliseconds with zero similar values,
the largest areas correspond to the attacks.
Additionally,
a time interval of \numprint{45000} milliseconds has been analysed to get a better view of the behaviour,
at the downside of fewer comparisons of motifs.
This leads to a higher false positive rate,
nicely shown for \textit{LIT-301} in Figure~\ref{fig:LIT-301_new}.
However,
the same consideration of \textit{DPIT-301} shows that the minimal distances are most prominent for the attacks,
presented in Figure~\ref{fig:DPIT-301_new}.
The prominent length of zeros for the similar motifs is presented as well.
From both these features,
information about anomalies due to attacks can be derived.


\subsection{\textit{Isolation Forest} and \textit{OCSVM}}
As described in Section~\ref{sec:algos},
\acp{ocsvm} and \textit{Isolation Forests} are one-class classifiers.
That means they are trained on one class of event while being capable of classifying between two classes in operation.
In this work,
the normal operation of the data set was employed as a training data set.
It contains \numprint{499220} instances of 51 sensor values each,
monitored during eleven days of operation.
After training,
the classifier was tested on the remaining four days of operation during which several attacks occurred.
The results are listed in Table~\ref{tab:ocsvm_res}.
\begin{table}[h!]
\renewcommand{\arraystretch}{1.3}
\caption{Performance of One-Class Classifiers}
\label{tab:ocsvm_res}
\centering
\scriptsize
\begin{tabular}{l l c r r c r r}
\toprule
\multicolumn{2}{c}{\textbf{\textit{Preprocessing}}} & \phantom{a} & \multicolumn{2}{c}{\textbf{\ac{ocsvm}}} & \phantom{a} & \multicolumn{2}{c}{\textbf{\textit{Isolation Forest}}} \\
& & & Accuracy & F1-Score & & Accuracy & F1-Score \\
\cmidrule{1-2} \cmidrule{4-5} \cmidrule{7-8}
\multirow{4}{*}{bool} & None & & \numprint{0.1214} & \numprint{0.2165} & & \numprint{0.2624} & \numprint{0.2058} \\
& 0 mean & & \numprint{0.1214} & \numprint{0.2617} & & \numprint{0.6182} & \numprint{0.1873} \\
& linear & & \numprint{0.2244} & \numprint{0.2125} & & \numprint{0.5762} & \numprint{0.1882} \\
& \textit{PCA} & & \numprint{0.3600} & \numprint{0.1995} & & \numprint{0.5514} & \numprint{0.1933} \\
\multirow{4}{*}{non\_bool} & None & & \numprint{0.1214} & \numprint{0.2165} & & \numprint{0.2660} & \numprint{0.2063} \\
& 0 mean & & \numprint{0.1214} & \numprint{0.2165} & & \numprint{0.4471} & \numprint{0.1953} \\
& linear & & \numprint{0.2409} & \numprint{0.2106} & & \numprint{0.4057} & \numprint{0.1968} \\
& \textit{PCA} & & \numprint{0.3511} & \numprint{0.1993} & & \numprint{0.4057} & \numprint{0.1992} \\
\bottomrule
\end{tabular}
\end{table}
In general,
the performance in these experiments is comparably bad.
In case of \textit{Isolation Forest},
pre-processing is able to improve the classification performance in terms of accuracy while reducing the F1-score.
In this case,
applying \ac{pca} procudes the best results.
\ac{pca} is a method to map multidimensional features to a lower dimensional feature space with the most important features being the most prominent in the output vector~\cite{Pearson.1901}.
The same goes for \ac{ocsvm},
with a smaller improvement because of pre-processing and a lower classification quality overall.
There are several hyper-parameters to be tuned.
However,
Changing $\nu$ and $\gamma$ for \ac{ocsvm} as well as changing the contamination factor or feature size for \textit{Isolation Forest} result in any significant improvement.
Furthermore,
the data has been evaluated with and without taking boolean values into consideration,
noted in Table~\ref{tab:ocsvm_res} as \textit{bool} and \textit{non\_bool}
preliminary experiments indicate that $\nu$ as a metric for training errors and support vectors does not have much influence on the result.
Additionally,
scaling the data before training and testing leads to only positive classifications in preliminary tests with a reduced data set.
Thus, 
all data in Table~\ref{tab:ocsvm_res} are derived from the unprocessed data set.
It should be noted that this classifier does not contain a notion of timing,
in contrast to the \textit{Matrix Profiles} and \acp{lstm}.

\section{Conclusion}
\label{sec:concl}
The digitisation of industry creates the demand for an increase in industrial cyber security solutions.
Due to legacy reasons,
they should integrate into existing applications.
The time series-based approach \textit{Matrix Profiles} is promising in terms of legacy-capabilities as well as detection.
Furthermore,
since little tuning of hyper-parameters is required, 
it is easy to set up and robust to different kinds of data.
Furthermore,
even though the detection capabilities are increased with a larger data base for comparison,
no formal training is required.
The addition to \textit{Matrix Profiles},
presented in this work,
can be used to detect attacks that occur multiple times and provide an increased level of security and traceability.
One-class classifiers on the other hand require extensive training with a large amount of data.
Despite pre-processing and tuning on hyper-parameters,
they do not perform satisfactorily for the data evaluated in this work.
Generally speaking,
the adaption of novel techniques for intrusion detection is necessary in order to meet the current requirements.
Combinations of anomaly detection methods with deception solutions,
such as presented by \textit{Fraunholz et al.}~\cite{Fraunholz.2017e, Fraunholz.2017f}.
Additionally,
presenting the results in a fashion that is easily understandable for human operators,
especially non-experts in cyber security is crucial for effective defense against attacks.
\textit{Lohfink et al.} present a visual representation of the results obtained in this work~\cite{Lohfink.2019}.

%
\begin{acks}
This work has been supported by the Federal Ministry of Education and Research of the Federal Republic of Germany (Foerderkennzeichen 16KIS0932, IUNO Insec) and the Deutsche Forschungsgemeinschaft (DFG,
German Research Foundation) – 252408385 – IRTG 2057.
The authors alone are responsible for the content of the paper.
\end{acks}

%
\bibliographystyle{ACM-Reference-Format}
\bibliography{literature}

\end{document}